\documentclass[prd,twocolumn,showpacs,preprintnumbers,amsmath,amssymb,aps,10pt]{revtex4-1}
\usepackage{graphicx}
\usepackage{dcolumn}
\usepackage{bm}
\usepackage{hyperref}
\usepackage{xcolor}
\usepackage{orcidlink}

\newcommand{\D}{\ensuremath{\mathrm{d}}}
\newcommand{\pdet}{\ensuremath{p_\mathrm{det}}}

\begin{document}


\title{Looking To The Horizon: Probing Evolution in the Black Hole Spectrum With Gravitational Wave Catalogs}
\author{Jam Sadiq\orcidlink{0000-0001-5931-3624}\textsuperscript{1, 2}}\thanks{Corresponding author, email: jamlucky.qau@gmail.com}
\author{Thomas Dent\orcidlink{0000-0003-1354-7809}\textsuperscript{3}}
\author{Ana Lorenzo-Medina\orcidlink{0009-0006-0860-5700}\textsuperscript{3}}

\affiliation{\textsuperscript{1} SISSA, Via Bonomea 265, 34136 Trieste, Italy and INFN Sezione di Trieste}
\affiliation{\textsuperscript{2} IFPU - Institute for Fundamental Physics of the Universe, Via Beirut 2, 34014 Trieste, Italy}
\affiliation{\textsuperscript{3} IGFAE, University of Santiago de Compostela, E-15782 Spain}

\date{\today}

\begin{abstract}
The population of black holes observed via gravitational waves currently covers the local universe up to a redshift $z\lesssim 1$, for the most massive merging binaries, or $z\lesssim 0.25$ for low-mass BH binaries (BBH).  Evolution of the BBH mass spectrum over cosmic time will be a significant probe of formation channels and environments. 
We demonstrate a reconstruction of the BBH merger rate, allowing for general dependence on binary masses and luminosity distance or redshift and accounting for selection effects, via iterative kernel density estimation (KDE) with optimized multidimensional bandwidths. 
Performing such reconstructions under a range of detailed assumptions, we see no significant evidence for the evolution of BBH masses with redshift, over the range where detected events are available.  
At most, possible trends towards increasing merger rate with redshift for primary masses $m_1\gtrsim 50\,M_\odot$, or towards decreasing merger rate with redshift for primary masses $m_1 \lesssim 40 M\odot$ may be supported. 
We compare these findings with previous investigations and caution against over-interpreting the current, sparse, data.  
Significantly upgraded detectors and/or facilities, and longer observing times, are required to harness any correlations of the BBH mass distribution with redshift. 
\end{abstract}


\maketitle

%
 \section{Introduction}

The detection of gravitational waves (GWs) \cite{LIGOScientific:2016aoc, LIGOScientific:2018mvr, LIGOScientific:2020ibl, LIGOScientific:2021usb, KAGRA:2021vkt} from merging binary black holes (BBHs) by advanced LIGO and Virgo and KAGRA~\cite{LIGOScientific:2014pky, VIRGO:2014yos, KAGRA:2020tym} has opened a new avenue for studying the astrophysical population of black holes \cite{LIGOScientific:2018jsj, LIGOScientific:2020kqk, KAGRA:2021duu}. The analysis of BBH populations using GW observations provides critical insights into their formation channels, mass distribution, spin properties, and their evolution across cosmic time. Gravitational wave observations have revealed that merger rates likely increase with redshift, aligning with the history of cosmic star formation~\cite{Madau:2014bja, Fishbach:2018edt, Callister:2020arv, KAGRA:2021duu, Schiebelbein-Zwack:2024roj}. Evidence has also emerged for characteristic features in the (primary) mass distribution, such as peaks near $10\,M_\odot$, and $35\,M_\odot$~\cite{Tiwari:2020otp,Sadiq:2021fin,Farah:2023vsc}, and that the majority of binaries display nearly equal mass ratios \cite{Talbot:2018cva, Tiwari:2021yvr, Farah:2023swu, Sadiq:2023zee}. 
Potential astrophysical correlations between source parameters 
have been analyzed to investigate binary formation channels; here we will focus on both the mass spectrum and evolution over cosmic time or equivalently over redshift~\cite{Fishbach:2018edt, Mapelli:2019bnp, KAGRA:2021duu, vanSon:2021zpk,Callister:2023tgi, Edelman:2022ydv,  Karathanasis:2022rtr, Rinaldi:2023bbd, Heinzel:2024hva, Lalleman:2025xcs}.

Despite these advances, the physical origins of these sources remain unclear. Numerous investigations have explored potential formation pathways \cite{Ivanova:2012vx, vandenHeuvel:2017pwp, Gallegos-Garcia:2021hti, Mandel:2015qlu, deMink:2016vkw, Marchant:2016wow, Downing:2009ag, Rodriguez:2015oxa, Rodriguez:2019huv, Mapelli:2021gyv, Miller:2008yw, Antonini:2016gqe, Mckernan:2017ssq, Stone:2016wzz, Silsbee:2016djf}, each predicting distinct gravitational wave signatures \cite{Bavera:2020uch, Bavera:2022mef, Zevin:2020gbd, Zevin:2022wrw, Broekgaarden:2022nst, Fuller:2019sxi, Bavera:2020inc, Fuller:2022ysb, Barkat:1967zz, Woosley:2016hmi, Tanikawa:2021zfm, Afroz:2024fzp}. Interpreting gravitational wave data within the context of these predictions should eventually yield astrophysical insights. 
 
One essential aspect of population analysis is understanding how the masses of BBHs correlate with redshift, which sheds light on their formation environments and underlying astrophysical processes. Recent studies have employed advanced statistical techniques to 
explore a possible correlation between BBH mass and redshift.
In \cite{Rinaldi:2023bbd}, a hierarchical Gaussian mixture is applied to changing properties of BBH over cosmic history, finding evidence of two distinct groups of black holes with different masses and redshifts, a lack of symmetric binary mergers, and a potential connection between black hole mass and the environment in which they formed. On the other hand \cite{Heinzel:2024hva} investigates the similar correlation using gravitational wave data with a method that avoids strong assumptions~\cite{Heinzel:2024jlc}, finding limited evidence for evolution in black hole masses over cosmic time, but possible connections between spin, mass ratio, and distance may exist. 
Non-parametric population reconstruction has also been proposed to investigate more general questions in GW cosmology and astrophysics~\cite{Ng:2024xps,Fabbri:2025faf}. 

Most recently, in \cite{Lalleman:2025xcs} the authors focus on the power-law distribution and peak features in the primary mass, 
carrying out a parameterized change-point analysis to investigate possible changes in such features over redshift: they find no significant evidence of evolution, although various types of variation in the mass spectrum cannot be excluded. 

This study investigates how the mass distribution of BBHs may evolve with redshift using publicly available gravitational wave observations. We employ an improved version of a non-parametric method based on iterative kernel density estimation (KDE), building on previous studies~\cite{Sadiq:2021fin,Sadiq:2023zee,Sadiq:2024xsz} in combination with an accurate fit of LVK search sensitivity~\cite{Lorenzo-Medina:2024opt}. 
Any evidence for such evolution would provide valuable insights into the astrophysical processes and formation channels that govern BBH systems. 
In~\cite{Sadiq:2021fin}, a preliminary study of the density of \emph{detected} BBH over mass and luminosity distance/redshift, using an earlier version of our adaptive KDE method but without accounting for selection effects, gave a qualitative indication that peak (local maximum) features in the mass distribution persisted over a range of distances.  Here, we revisit that initial hint with several methodological improvements, including a self-consistent treatment of event measurement uncertainties~\cite{Sadiq:2023zee} and, crucially, an accurate estimate of the selection function over binary masses and distance introduced in~\cite{Lorenzo-Medina:2024opt} that enables us to probe the astrophysical rate density.

A critical point given the current small statistics of GW detections is how far statistical inferences and scientific conclusions depend on model assumptions; and conversely, how sensitive they are to fluctuations in the observed GW event set, considered as a random sample.  Clearly, little information can be obtained 
in regions of parameter space where no events are detected. 

The most obvious statistical limitation is the selection function restricting the visibility of BBH mergers at finite distance, due to detector and search sensitivity: the selection function is also strongly dependent on binary component masses (see e.g.\ discussion in~\cite{KAGRA:2021vkt,Fishbach:2017zga}, and \cite{Lorenzo-Medina:2024opt}).  Thus, the BBH mass spectrum around $\sim\!10\,M_\odot$ can only be investigated out to fairly small redshift ($z\lesssim 0.25$) with current data. 
However even at higher masses, for which the current horizon of detectability extends farther, our knowledge of the BBH population is limited simply by the scarcity of mergers.  We already detect a high fraction of high-mass BBH at low redshift, but this is still a very small number of events: thus, only significantly longer observing times will eventually enable an accurate local measurement. 

Issues of selection bias and finite number statistics were explored in a study of a strongly redshift-dependent mock data population of ``light seed'' black holes observable by LISA, using the same KDE-based method, in~\cite{Sadiq:2024xsz}.  There, we found that the true population could be reconstructed by the method, but only within parameter regions with sufficient numbers of detections.
For BBH currently observed by LIGO-Virgo, our knowledge of the selection function, as well as inspection of detected events, indicate that uncertainties in any current measurement of the redshift-dependent mass function are necessarily large. Thus, nontrivial model assumptions would probably be necessary to obtain any significant evidence of evolution.

The remainder of this paper is structured as follows: in section \ref{sec:method_analysis} we describe our method and in section \ref{sec:application} its application to the GWTC-3 data in this study. In section \ref{sec:results} we present results from this data, and discuss the implications for astrophysics and future GW observations in section~\ref{sec:discussion}.

\section{Population reconstruction with iterative reweighted KDE}
\label{sec:method_analysis}

In \cite{Sadiq:2021fin}, we proposed adaptive Kernel Density Estimation (KDE) as a non-parametric method to study the distribution of BH primary masses using GW observations. In \cite{Sadiq:2023zee}, we improved this reconstruction method by using an iterative reweighting technique to account for parameter measurement uncertainties and the selection function, and also extended its application to the two-dimensional parameter space of BBH component masses.  Here, we will briefly recap the main features of adaptive KDE and describe the improvement applied in this work, specifically the use of optimized non-isotropic kernels, as opposed to the isotropic (spherical) kernels used in previous studies. 

\subsection{\textbf{Summary of previous work}}

The estimated probability density $\hat{f}(\vec{x})$ at point $\vec{x}$ from a multidimensional Gaussian KDE is 
\begin{multline}
    \hat{f}(\vec{x}) = \frac{1}{N \sqrt{(2\pi)^D|\Sigma|}} \sum_{i=1}^N \frac{1}{\lambda_i} \cdot \\
    \exp\left(-\frac{1}{2} (\vec{x} - \vec{X}_i)^T \lambda_i^{-2} \Sigma ^{-1} (\vec{x} - \vec{X_i})\right)
\end{multline}
where $i = 1 \ldots N$ labels the observations $\vec{X}_i$, $D$ is the dimensionality of the data, $\Sigma$ 
is a global kernel covariance matrix corresponding to the KDE bandwidth, and the local adaptive parameter $\lambda_i$ is given by  
\begin{equation}
 \lambda_i = \left( \frac{\hat{f}_0(\vec{X}_i)}{g} \right)^{-\alpha}, \, \, \log g = N^{-1} \sum_{i=1}^{N} \log \hat{f}_0(\vec{X}_i)\,.
\end{equation}
Here $\hat{f}_0$ is an initial pilot density estimate obtained by setting $\lambda_i=1$ and $\alpha$ is the bandwidth sensitivity parameter (lying between $0$ and $1$), $g$ being a normalization factor. The optimized choices of 
the kernel covariance (i.e.\ bandwidth) and $\alpha$ are determined by maximizing a likelihood figure of merit, evaluated via cross-validation~\cite{Sadiq:2021fin}: 
\begin{equation}
  \log \mathcal{L}_{\rm Kfold} = \sum_{k=1}^{K} \sum_{i \in \text{Fold}_k} \ln \hat{f}_{{\rm Fold}-k}(X_i).
\end{equation} 
Here the data is partitioned into $K$ `folds' of equal size; $\hat{f}_{{\rm Fold}-k}$ is the KDE trained on all folds \emph{except} $\text{Fold}_k$.  The log-likelihood score penalizes relative errors, making it 
suitable for estimation across a wide dynamic range of densities.

As we use \emph{detected} events to optimize and evaluate the KDE, reconstruction of the actual astrophysical event density also requires an estimate of the selection function, as discussed below in Sec.~\ref{ssec:pdet}. 

In \cite{Sadiq:2023zee}, we improved the accuracy of our method by employing an iterative approach similar to Expectation-Maximization (EM) algorithms.  The uncertainty in measured individual event properties, which for GW detections is often large, is represented by parameter estimation (PE) samples obtained by Bayesian inference (see e.g.~\cite{Veitch:2014wba}) using a standard default prior over mass and luminosity distance.  Since these priors do not reflect the actual astrophysical population, the resulting parameter estimates are necessarily biased.  Our iterative reweighting method is designed to correct for this bias self-consistently: we reweight the samples for each event at a given iterative step using density estimates from the previous step, adjusted for the selection function.  We expect these estimates to, on average, have decreasing bias with successive iterations, as the KDE likewise more closely approaches the true (detected) distribution.

At every reweighting step, we draw Poisson (1) samples for each detected event~\cite{Sadiq:2023zee}, without replacement in the case of $>1$ samples, to avoid corner cases in cross-validation where the same sample point could appear in different folds.   
As with Markov Chain Monte Carlo methods, we desire to reach a steady state where successive iterations are fair draws from a distribution of density estimates representing the uncertainties on a reconstructed population: both due to finite event statistics and to individual event measurement errors. 

We monitor the autocorrelation of the optimized bandwidth and adaptive parameter over iterations as diagnostics for the approach to stationarity.  We discard an initial burn-in phase, and then establish a buffer of 100 iterations from which we compute a mean KDE; finally we obtain $500$ or more independent population estimates by bootstrap resampling as above, but using this buffer mean KDE to determine the reweighting of PE samples (in place of the previous Markov chain iteration).  
We quantify the population distribution via the median and 5th and 95th percentiles of these independent estimates.

This method is a computationally efficient approach to obtain self-consistent density and rate estimates, including bootstrap uncertainties.  However, the method as previously applied is limited by the use of spherical (isotropic) kernels, implying $\Sigma = h^2\mathbb{I}$ in standardized coordinates, where $\mathbb{I}$ represents the identity matrix.  This restriction can introduce bias in the estimates when the data follow distinct distributions, or have significantly different measurement uncertainties, over different dimensions. 

\subsection{\textbf{Improvement to this work}}

Here we address this limitation 
by allowing for kernels with different relative bandwidths along each dimension, while the overall kernel scale varies between data points as already described for adaptive KDE. 
For a data point $\vec{X}_i$, the kernel contribution is proportional to
\[
\exp\left[ -\frac{1}{2} (\vec{x} - \vec{X}_i)^T \lambda_i^{-2} \mathrm{diag}(h_1^{-2}, h_2^{-2}, \ldots) (\vec{x} - \vec{X}_i) \right],
\]
where $h_1, \ldots h_D$ are the global $D$-dimensional bandwidths and $\lambda_i$ is the per-point adaptive factor as before.
We implement this general multi-dimensional KDE using KDEpy~\cite{tommy_odland_2018_2392268}
via rescaling the data. 
For coordinates $\vec{x}$ where the data is standardized to unit variance in each dimension, the rescaling is defined as
$\vec{y} = \left( h_1^{-1} x_1, h_2^{-1} x_2, \ldots \right)^T,
$
and the kernel in rescaled coordinates becomes
\[
\exp\left[ -\frac{1}{2} (\vec{y} - \vec{Y}_i)^T \lambda_i^{-2} \mathbb{I} (\vec{y} - \vec{Y}_i) \right].
\]
This is not the most general linear data transformation, which would be $\vec{y} = H^{-1}\vec{x}$ where $H^{-1}$ is any invertible real matrix, corresponding to an off-diagonal kernel bandwidth matrix.  We choose not to pursue this more complicated option: such kernels might increase apparent correlations between parameters, thus given our choice of diagonal kernel we expect our conclusions on correlations or evolution to be conservative. 

The $D$-dimensional bandwidths $h_1, \ldots h_D$ 
need to be optimized, along with the adaptive parameter $\alpha$. 
Given the dimensionality of this hyper-parameter space, instead of the grid search used in previous works we switch to a generalized optimization algorithm, in this case Nelder-Mead. Our cross-validated likelihood figure of merit is not invariant under rescaling coordinates: to ensure stable optimization we evaluate it over the $\vec{x}$ coordinates, i.e.\ without rescaling~\footnote{Standardization also changes the likelihood figure of merit, but only by a constant for a given data set.}.

\begin{figure}[tbp]
    \centering
        \includegraphics[width=0.8\linewidth]{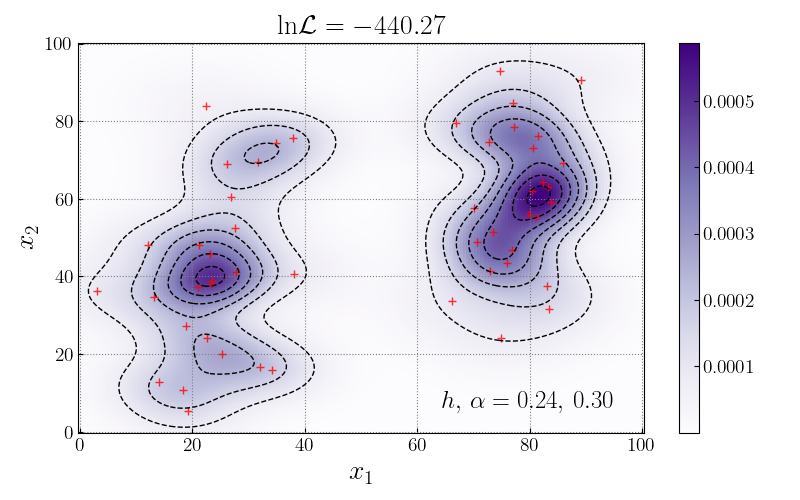} \\
        \includegraphics[width=0.8\linewidth]{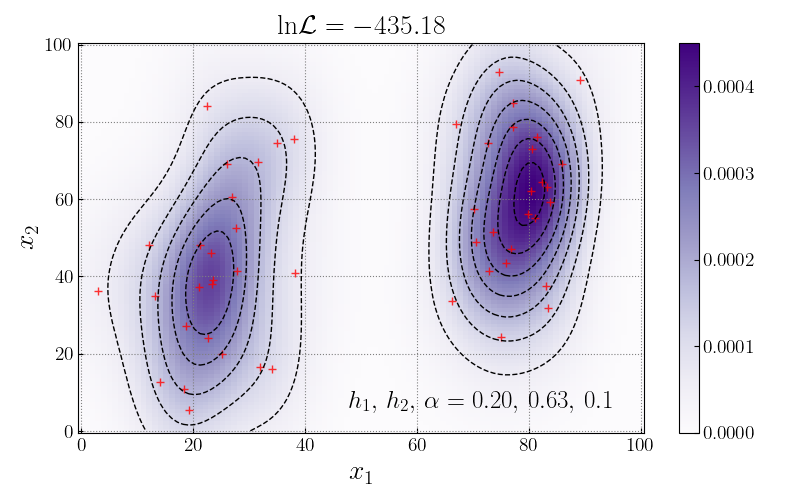}
    \caption{Verification of the multidimensional bandwidth optimization using mock data consisting of a mixture of two Gaussian components, represented by red $+$ symbols: contours and blue shading show the KDE in each case. 
    Top: Adaptive KDE with an isotropic kernel (in standardized coordinates): the single bandwidth parameter $h$ and adaptive parameter $\alpha$ optimize the cross-validated likelihood.  
    Bottom: Adaptive KDE with different optimized bandwidths $h_{1,2}$ for each coordinate. 
    }
    \label{fig:mock-data-comparison}
\end{figure}
We validate the multi-dimensional kernel optimization 
using synthetic data comprising a bimodal Gaussian mixture. This dataset is useful for testing because the isotropic (spherical) kernel lacks the flexibility to capture its multimodality by using different bandwidths per parameter, even after standardizing the data. 

We first apply adaptive KDE with an isotropic kernel in standardized coordinates, optimizing a single bandwidth parameter $h$ and adaptive parameter $\alpha$ to maximize the likelihood via leave-one-out cross-validation. However the resulting estimate, shown on the left of Fig.~\ref{fig:mock-data-comparison}, is under-smoothed along one of the two dimensions, introducing excess variance and failing to represent the true distribution.  Our improved KDE with independently optimized bandwidths $h_1$ and $h_2$, shown in Fig.~\ref{fig:mock-data-comparison} (right), 
better accommodates the anisotropic nature of the data, successfully recovering the bimodal Gaussian mixture, as well as significantly increasing the likelihood.

\section{Application to GWTC-3 data}
\label{sec:application}

We apply the reweighted adaptive KDE with multi-dimensional bandwidth optimization to the results from parameter estimation \cite{ligo_scientific_collaboration_and_virgo_2021_5546663} of gravitational-wave observations cataloged in GWTC-3 \cite{KAGRA:2021vkt, KAGRA:2023pio}. 

\subsection{Event selection, parameter estimation and data cleaning}
Following the approach in our earlier studies \cite{Sadiq:2021fin,Sadiq:2023zee}, we selected 69 BBH candidate events from GWTC-3 with false alarm rates below 1 per year, excluding GW190814, an outlier event with an unusually low secondary mass ($m_2\simeq2.6,M_\odot$, $q\simeq0.11$), as its nature remains uncertain: either a very massive neutron star or a light black hole, which is inconsistent with the \textbf{bulk} BBH population we aim to analyze \cite{LIGOScientific:2020zkf}. 

In LIGO-Virgo-KAGRA (LVK) catalog PE analyses, the prior over masses and $d_L$ is uniform over redshifted masses $m_{\{1,2\}z} = m_{\{1,2\}} (1 + z)$ and over Euclidean volume~\cite{LIGOScientific:2020ibl}, thus is proportional to $(1+z)^2$ $d_L^2 \D m_1 \D m_2 \D d_L$ (e.g.~\cite{Callister:2021gxf}).  Samples reweighted to correspond to a ``cosmological'' prior uniform over comoving volume and source-frame time, though still uniform over $m_z$, are also available and are the basis of quoted event parameters~\cite{KAGRA:2021vkt}.  We use the original samples and apply a factor proportional to the inverse prior $(1+z)^{-2} d_L^{-2}$ to our estimate of $p(m_1[,m_2],d_L)$ in the reweighting step of our algorithm. 

For two events with relatively low SNR, GW190719\_215514 and GW190805\_211137, we find a small number of PE samples at high distance which appear to represent the prior distribution, which is formally divergent towards large distance, rather than estimating the parameters of the detected merger.  The expected (optimal) SNRs for these samples are unusually small. We compare the distances of PE samples with those of simulated signals (injections) which are subject to an expected network SNR cut of 6, as lower SNRs are considered ``hopelessly'' undetectable~\cite{ligo_scientific_collaboration_and_virgo_2023_7890437,ligo_scientific_collaboration_and_virgo_2021_5636816}.  We exclude samples at higher distance than any injection with comparable masses (similar data cleaning was performed in \cite{Sadiq:2024xsz}).

\subsection{Selection effects}
\label{ssec:pdet}

For the search sensitivity estimate or selection function, we employ an analytic approximant~\cite{Lorenzo-Medina:2024opt} which accurately fits the probability of detection $\pdet$ at the 1 per year FAR threshold for injection (simulated signal) results released with the catalog \cite{ligo_scientific_collaboration_and_virgo_2021_5636816}.  
The $\pdet$ approximant is a function of masses, effective inspiral spin $\chi_{\rm eff}$ \cite{Ajith:2009bn} and luminosity distance $d_L$, marginalizing over the sky direction, source orientation and spin orientations at given $\chi_{\rm eff}$. 
For our main results, we approximate $\chi_\mathrm{eff}$ to zero when evaluating the selection function for simplicity, motivated by findings in \cite{KAGRA:2021duu} which support small effective spins for the observed BBH population. 

\begin{figure}[tbp]
    \includegraphics[width=0.9\linewidth]{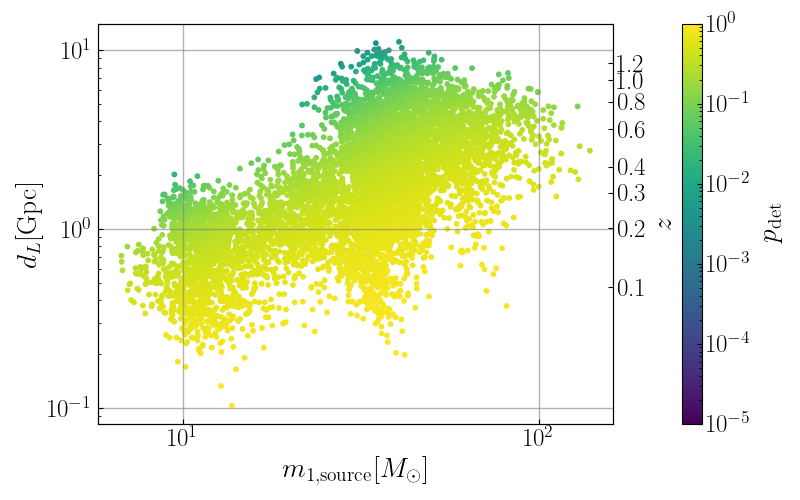}
    \includegraphics[width=0.9\linewidth]{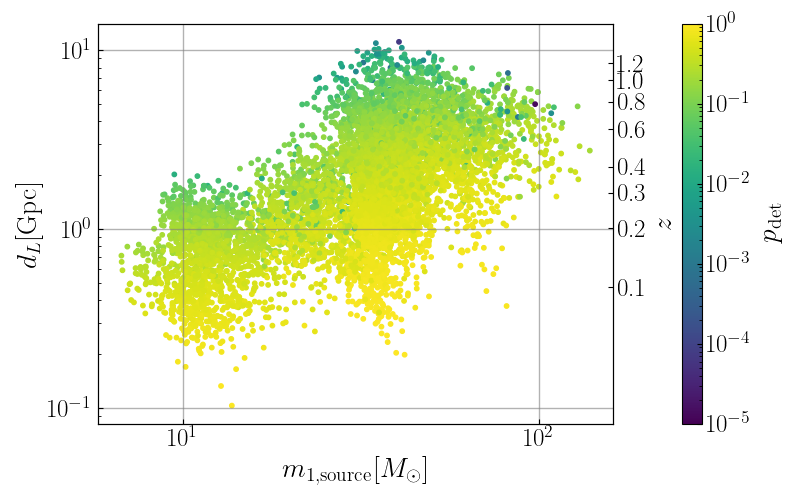}
    \caption{Detection probability $\pdet$ as a function of the primary mass $m_1$ and luminosity distance $d_L$ for PE samples from 69 BBH events in GWTC-3. The top panel assumes a power-law distribution for the mass ratio, $p(q) \propto q^{1.26}$, while the bottom panel uses posterior sample values of secondary mass $m_2$.
    }
    \label{fig:pdet-on-pesamples}
\end{figure}

We perform two separate analyses over masses and distance. In the first analysis, we reconstruct the population over primary mass and distance while assuming a power-law distribution of mass ratio $q = m_2/m_1$: here we compute $\pdet$ as a function of $m_1$ and $d_L$, while marginalizing over the secondary mass assuming a power-law distribution $\sim q^{\beta}$, see Fig.~\ref{fig:pdet-on-pesamples} (top), using the median estimate $\beta = 1.26$ from the \textsc{Truncated} analysis of~\cite{LIGOScientific:2020kqk}, as a simple initial assumption. 

In the second analysis, we reconstruct the 3d population over both component masses and distance: hence we evaluate $\pdet$ as a function of $m_1$, $m_2$, and $d_L$, see Fig.~\ref{fig:pdet-on-pesamples} (bottom). 
%
We conduct an additional check on the approximation of $\chi_{\rm eff}$ to zero by computing $\pdet$ based on the $\chi_\mathrm{eff}$ value of PE samples; see Fig.~\ref{fig:Xieffbasedpdet-on-pesamples} for a direct comparison.  Corresponding results 
are provided in Appendix~\ref{app:xieffpdet}. 

The fitted detection probability is a function of the redshifted masses $(1+z)m_{1,2}$ which describe the GW signal at the detectors.  
To relate the redshift to $d_L$ we assume a standard \textit{flat $\Lambda$CDM} cosmology with $H_0 = 67.9\,$km/s/Mpc and $\Omega_M = 0.3065$ \cite{Planck:2015fie}, as also used for LVK parameter estimation; changes in the assumed cosmology, particularly in $H_0$, would have minor effects on the reconstructed source mass spectrum. 

In some regions of the parameter space, principally at high distance, $\pdet$ can become extremely small. Our KDE of detected events will also take low values in regions with small numbers of samples. In principle, our estimate of the astrophysical merger density is proportional to the ratio of the KDE to $\pdet$; however, in the limit of very small $\pdet$ this estimate will become numerically unstable.  To mitigate or regularize this issue, we limit the range of $\pdet$ used in the analysis by replacing (capping) values below a predefined threshold $\pdet=0.1$ by the threshold value, as demonstrated in the context of LISA mock data in \cite{Sadiq:2024xsz}. 
%
As shown in Fig.~\ref{fig:pdet-on-pesamples}, most samples have $\pdet$ values above or around $0.1$, but a small fraction have significantly lower values, 
particularly at larger distances or for higher primary masses.  
As a result of this capping or regularization we expect the rate estimate to be biased downwards, and eventually to approach zero, in regions with very small $\pdet$.  In any case no useful information is available about the rate density in such ``censored'' regions, as they were dubbed in \cite{Rinaldi:2023bbd}.

\subsection{Limits on bandwidths}
\label{ssec:bw}

As described in Section \ref{sec:method_analysis}, we optimize the KDE hyperparameters using $k$-fold cross-validation with a likelihood figure of merit. We employ the Nelder-Mead optimizer to determine the optimal bandwidths $h_k$ for each dimension and the adaptive parameter $\alpha$: by default the allowed bandwidths for each dimension range between 0.01 and 1 (when considering standardized data). 

In the two-dimensional analysis where a power-law assumption is applied to 
mass ratio, 
the distribution of optimized bandwidths over our bootstrap iterations is approximately Gaussian for both $m_1$ and $d_L$, and no iterations produce bandwidths smaller than 0.1.

However, in the three-dimensional analysis 
the optimized bandwidths for the $d_L$ dimension drop to $\lesssim\!0.1$ for a fraction of iterations. This behavior likely arises due to random clustering of samples at nearby $d_L$ values (see Appendix~\ref{app:bwdLsmaller}). 
Density estimates for such small $d_L$ bandwidth values show several separate local maxima over distance or redshift; however, as this implies a large non-monotonic variation of the merger rate, we consider such behaviour astrophysically implausible.  In addition, such wide rate variation contributes to excess variance in the iterative estimate. 

To address this, in the 3d analysis we restrict the minimum bandwidth in the $d_L$ dimension (taking standardized data) to $0.3$, such that multimodal distributions over $d_L$ are largely eliminated. 
This has the potential drawback that the analysis would be unable to reconstruct rapid changes in the rate over distance or redshift. 
Thus, for completeness, we will provide results of the three-dimensional analysis without restrictions on the $d_L$ bandwidth in Appendix~\ref{app:bwdLsmaller}. 

\section{Results}
\label{sec:results}

\subsection{2d primary mass vs.\ distance}

We first apply our adaptive KDE with optimized non-isotropic kernels to reconstruct the BBH merger rate dependence on the primary mass and luminosity distance, assuming a power-law distribution over mass ratio when computing selection effects. 
Our KDE is implemented over $\log m_1$ and $d_L$: i.e.,\ the Gaussian kernel takes a constant form specified by bandwidths $h_{m_1}$, $h_{d_L}$ in these coordinates.  The resulting KDE is transformed to obtain a normalized density over $m_1$ and $d_L$.  We then infer the rate density of arrival of GW events over this space by dividing by the estimated detection probability as in \ref{ssec:pdet} and multiplying by the number of detected events per observation time.  We finally convert this KDE rate density to a cosmological merger rate density in units of comoving volume and source-frame time, as detailed in Appendix~\ref{app:comov_rate}.

The analysis uses 1100 iterative reweighting steps, where the first 100 iterations form a ``burn-in'' period and are discarded.
\begin{figure}[tbp]
    \includegraphics[width=0.95\linewidth]{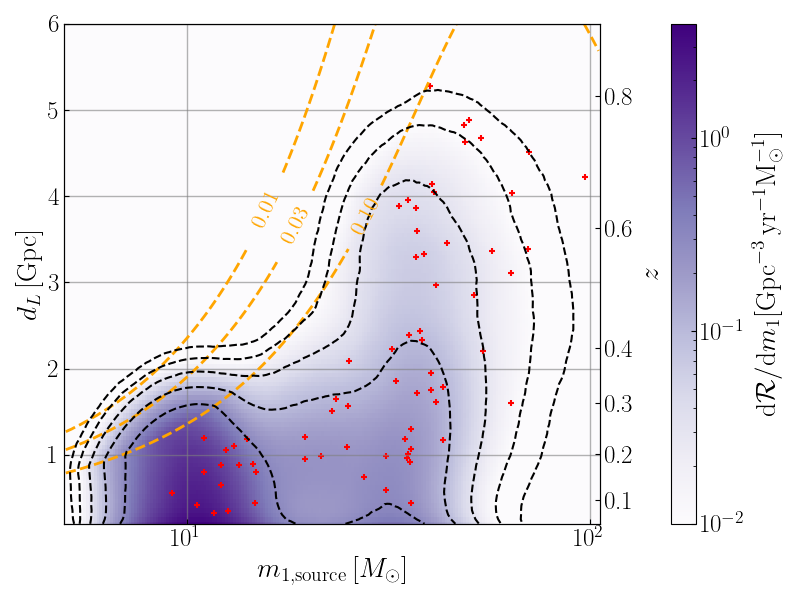}
    \caption{
    Rate density as a function of primary mass $m_1$ and luminosity distance $d_L$, assuming a power-law distribution for $m_2$. Red \textbf{$+$ symbols} represent the medians of PE samples \textbf{for GWTC-3 BBH events}, while the orange contours correspond to $\pdet$ levels. One event, GW190805\_211137, has a median $d_L \sim 7500$\,Mpc outside the range of the plot.
    }
    \label{fig:m1dL2D_withselectioneffectContours}
\end{figure}
We plot the median rate density of the remaining 1000 iterations in Fig.~\ref{fig:m1dL2D_withselectioneffectContours}. The median PE samples for all events lie within the $\pdet>0.1$ contour except one, GW190805\_211137, with a median distance $\sim\! 7500$\,Mpc outside the plotted range.  Thus, we may only expect an accurate or informative rate estimate \emph{within} the $\pdet > 0.1$ contour.  For some ranges of primary mass the rate estimate may not even be valid up to this point, simply due to a low count of detected events: for instance for $m_1 \lesssim 9\,M_\odot$ there is only a single detection, hence our rate estimate is effectively an extrapolation from higher mass events. 

We observe a clear maximum around $m_1\simeq 10\,M_\odot$ which persists with approximately constant rate density up to the limit of validity of our reconstruction, $d_L\simeq 1.2\,$Gpc for this mass range. 
An underdensity just below $\sim\!20\,M_\odot$ again persists to the limit $\pdet \lesssim 0.1$. 
A second overdensity at roughly $m_1\simeq 35\,M_\odot$ again persists to the limit of detected events around $\sim\!4\,$Gpc, although with minor apparent variation in rate. 
Finally, the rate falls off at high mass $m_2\gtrsim 50\,M_\odot$: this falloff appears to be more rapid at low distance / redshift, with more support at distance $\sim\!3\,$Gpc, hinting at possible evolution of the high-mass end of the population. 

To display any evolution over redshift, we 
extract the rate density over $m_1$ at a sequence of $d_L$ or redshift values.
\begin{figure}[tbp]
    \hspace*{0.0cm}
    \includegraphics[width=0.93\linewidth]{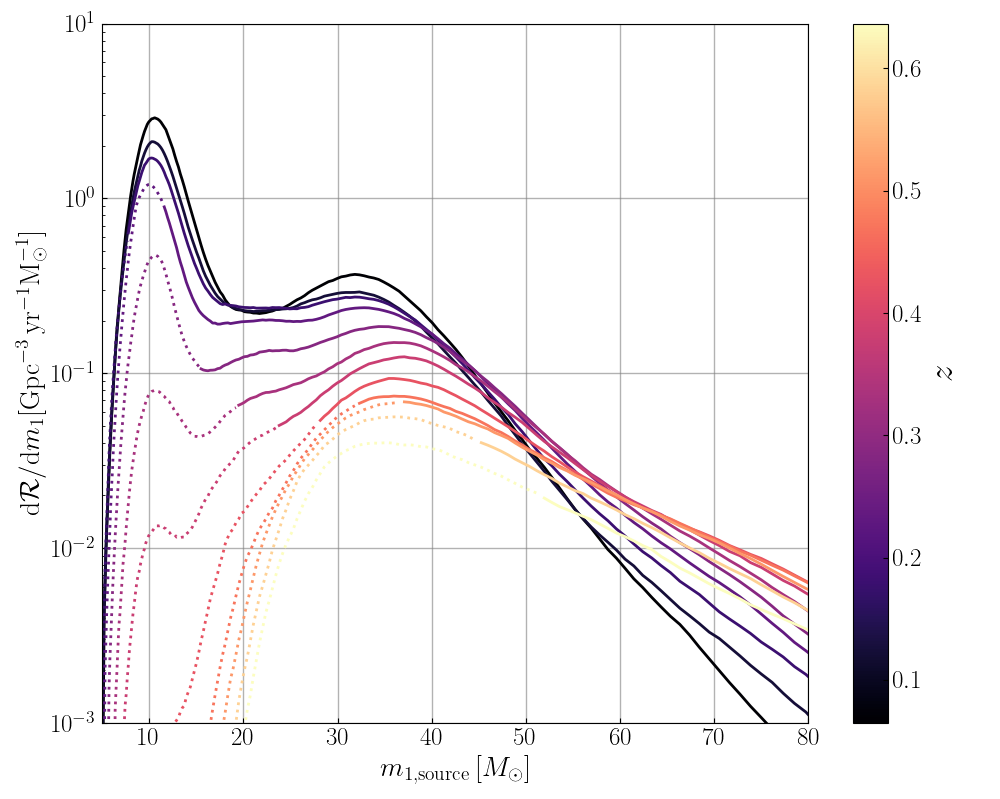} \\
    \hspace*{-0.6cm}
     \includegraphics[width=0.83\linewidth]{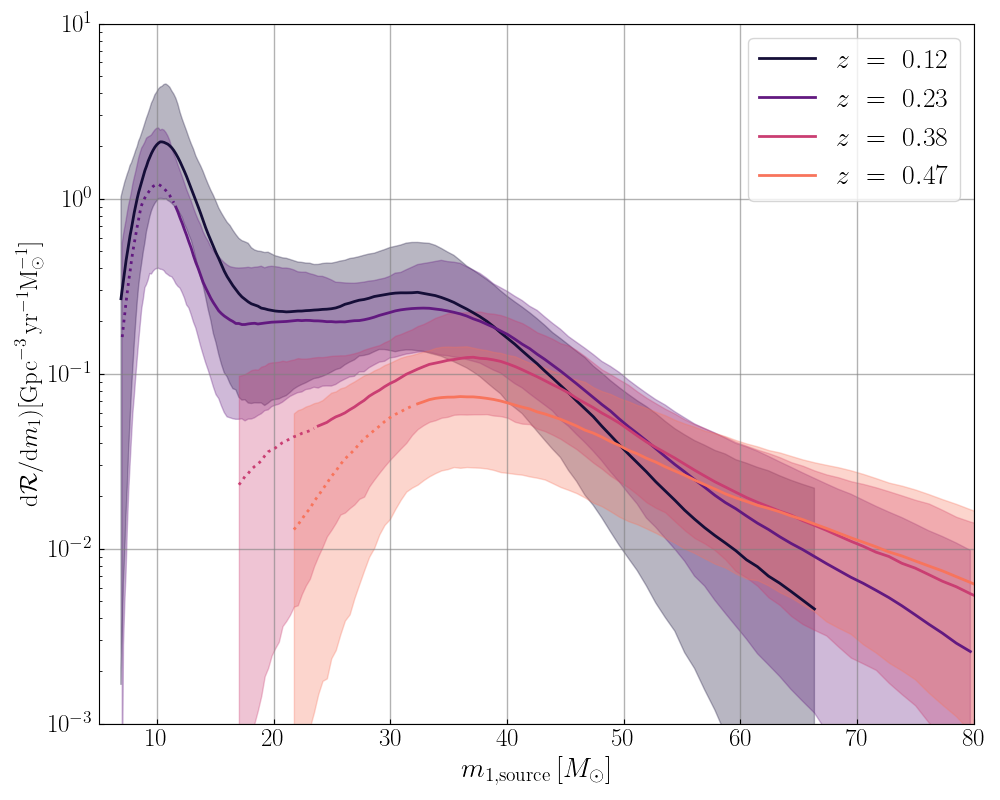}
    \caption{Top panel: Merger rate density (median estimate) as a function of the primary mass $m_1$, with different curves representing redshift values indicated by color, assuming a power law distribution of mass ratio.
    Estimates at low mass are increasingly affected by censorship towards high redshift: corresponding regions with $\pdet<0.1$ are shown as dotted curves.  Bottom panel: Rate density for a subset of redshift values, 
    with shaded 90\% uncertainty intervals.  We omit values with excessively high uncertainty for clarity (as also in Fig.~\ref{fig:offset-m1-rate-powerlaw-m2}).
    }
    \label{fig:median-m1-rate-powerlaw-m2}
\end{figure}
We show median rate values in Fig.~\ref{fig:median-m1-rate-powerlaw-m2}, with 90\% uncertainty bands for selected redshifts (omitting points with extremely large uncertainty), and show uncertainties for all redshifts in Fig.~\ref{fig:offset-m1-rate-powerlaw-m2}. 
\begin{figure}[tbp]
    \centering
    \includegraphics[width=0.99\linewidth]{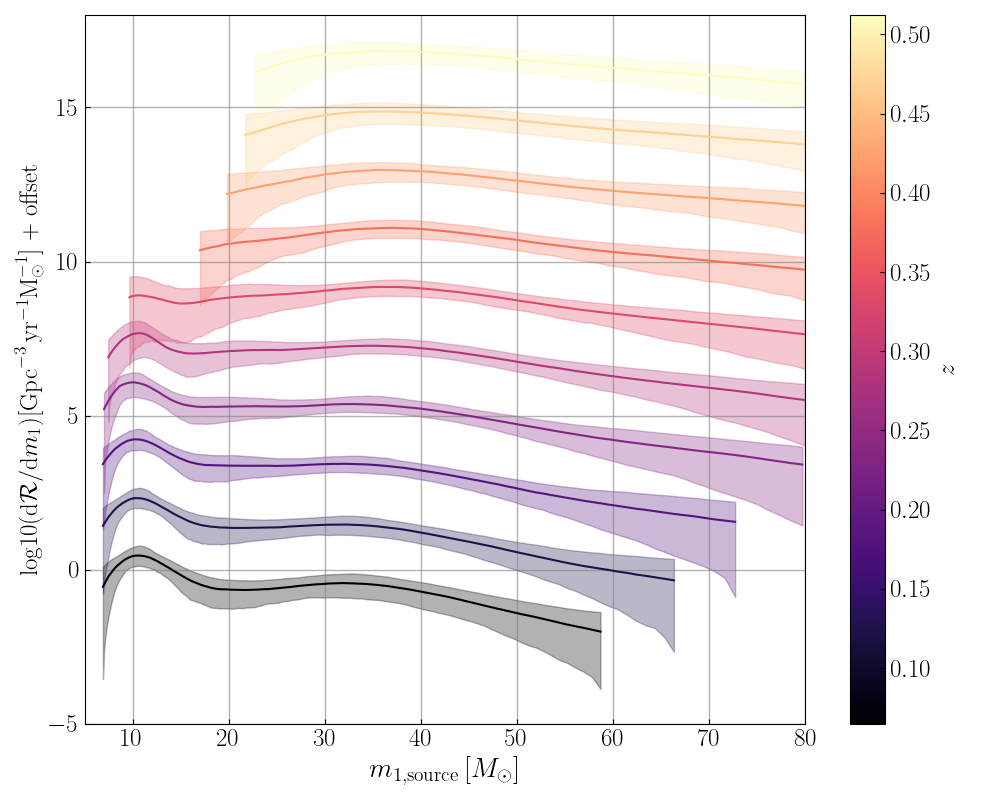}
    \caption{Merger rate density as a function of primary mass with symmetric 90\% confidence intervals, corresponding to the results shown in Fig.~\ref{fig:median-m1-rate-powerlaw-m2}. 
    Each curve shows a constant distance or redshift, shown by the color scale, with constant offsets applied for clarity. We only plot values where both (1) the 5th percentile is $>10^{-3}\times$ median, and (2) the 95th percentile is $\leq 200\times$ the median.
    }
    \label{fig:offset-m1-rate-powerlaw-m2}
\end{figure}

We observe two main apparent trends in Fig.~\ref{fig:median-m1-rate-powerlaw-m2}: first, the height of the $10\,M_\odot$ peak, and the distribution below $30\,M_\odot$ as a whole, decreases rapidly at higher redshift; however this can be largely attributed to ``censorship'' in this region, i.e.\ absence of detected events as expected for very small detection probability.  At intermediate masses ($m_1 \sim 35 \,M_\odot$) there is a weak apparent trend toward decreasing rate with redshift.  However the uncertainty bands in the lower plot still overlap for mass ranges unaffected by censorship, hence we do not find the trend to be significant.
Second, the falloff at high mass ($>\!50\,M_\odot$) becomes shallower at higher redshift.  Although this region is fully accessible to the detector network, Fig.~\ref{fig:offset-m1-rate-powerlaw-m2} indicates that rate uncertainties are very large, as expected given the low statistics of detected high-mass events.  Overall, this analysis is consistent with no evolution of the primary mass distribution \textbf{given uncertainties due to finite statistics}. 

\subsection{3d component masses vs.\ distance}

We extend the analysis to three dimensions, imposing a constraint on the bandwidth in the $d_L$ dimension to be $\geq 0.3$ as discussed in \ref{ssec:bw}. 
The 3D analysis, as compared to the 2D case, removes the assumption of a power-law distribution for the mass ratio $m_2/m_1$ in selection effects, and allows us to investigate variations in the component mass distribution as a whole, including correlations.
Analogously to previous cases we take a Gaussian kernel over $(\log m_1, \log m_2, d_L)$. 
\begin{figure}[tbp]
    \includegraphics[width=0.9\linewidth]{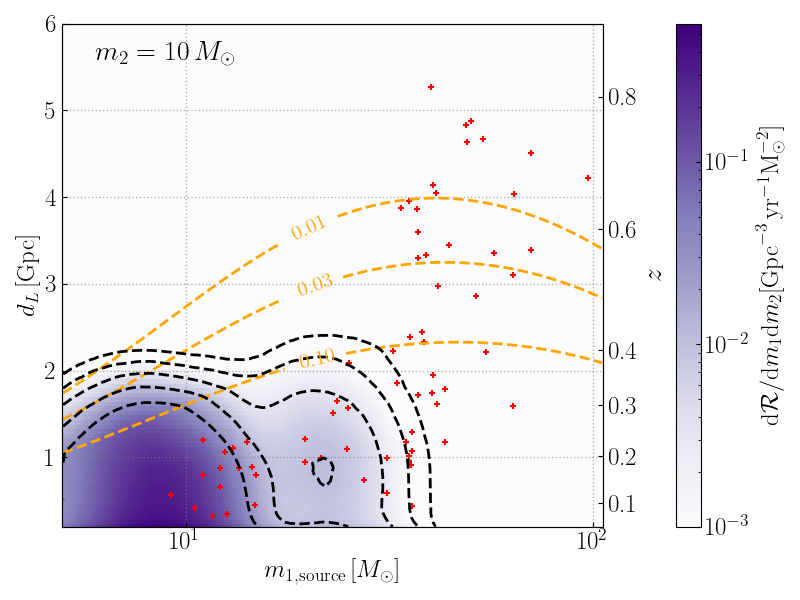} 
    \includegraphics[width=0.9\linewidth]{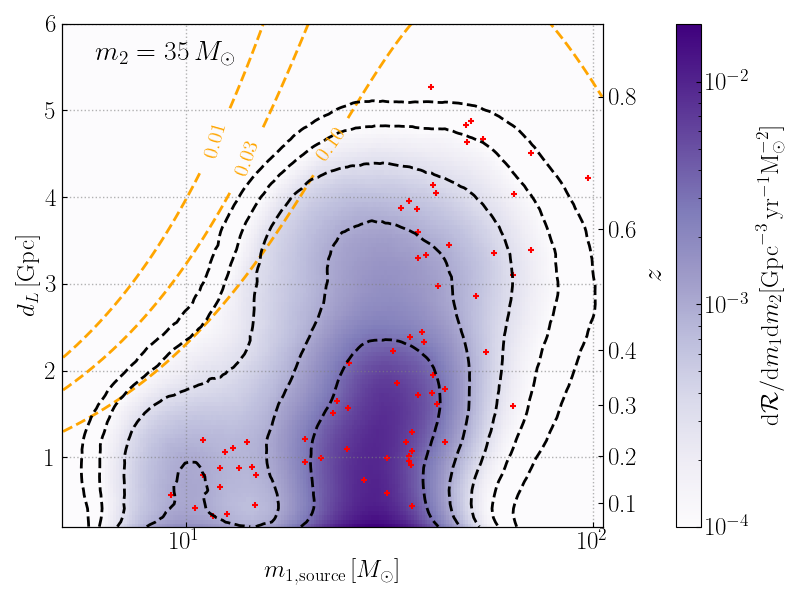}
    \caption{BBH rate density as a function of component mass $m_1$ and luminosity distance $d_L$, for fixed $m_2$.
    Note that, for this plot only, we do not impose $m_1>m_2$ in displaying the KDE. 
    The orange contours represent $\pdet$ 
    at the given $m_2$ values. Red $+$ markers indicate medians of PE samples. 
    }
    \label{fig:fixedm2slice-3Dcase-m1dLrates}
\end{figure}

From the 3D results, we compute rates for fixed values of $m_2$, as shown in Fig.~\ref{fig:fixedm2slice-3Dcase-m1dLrates} for $m_2=(10, 35)\,M_\odot$, overplotting contours of $\pdet$. 
(Here, unlike the rest of our results, we do not impose $m_2<m_1$: thus these plots show a slice through the full 2-d component mass plane.) 
For $m_2=10\,M_\odot$, the overdensity around $m_1\simeq 10\,M_\odot$ persists up to the limit of detected events around $1\,$Gpc ($z\simeq 0.2$). 
For $m_2=35\,M_\odot$, the clear peak at $m_1\simeq 35\,M_\odot$ persists out to a distance $\sim 4\,$Gpc ($z\simeq 0.6$), however detected events fall somewhat short of the $\pdet=0.1$ contour, indicating that caution should be used when interpreting the reconstruction at such large distances.  These features are roughly as expected from the 2D results shown in Fig.~\ref{fig:m1dL2D_withselectioneffectContours}. 
Our comoving rate estimates also show an apparent increase at very low distance $d_L\lesssim 0.3\,$Gpc: this is an artefact of our choice to construct a KDE over $d_L$, as the kernels have nonzero support at zero distance.  Hence, we do not expect the estimate to be valid at distances below the closest detected events ($\sim\! 0.3\,$Gpc).


\begin{figure}[tbp]
\centering
\includegraphics[width=0.9\linewidth]{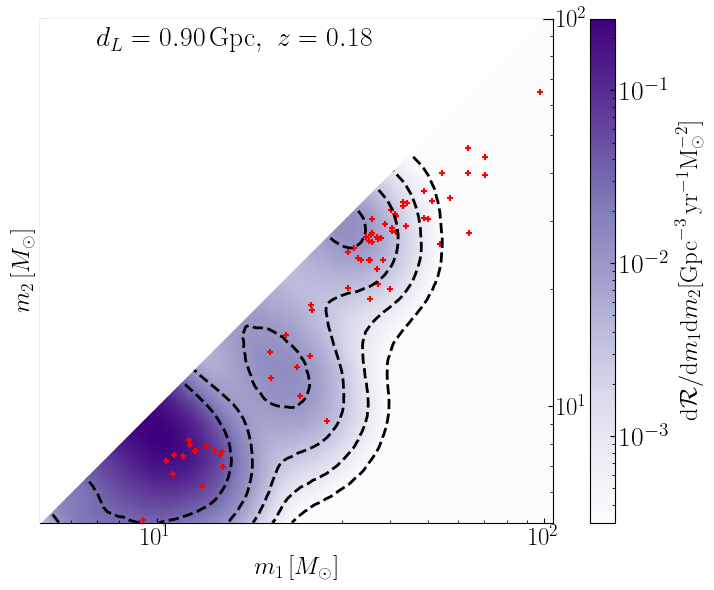}
\includegraphics[width=0.9\linewidth]{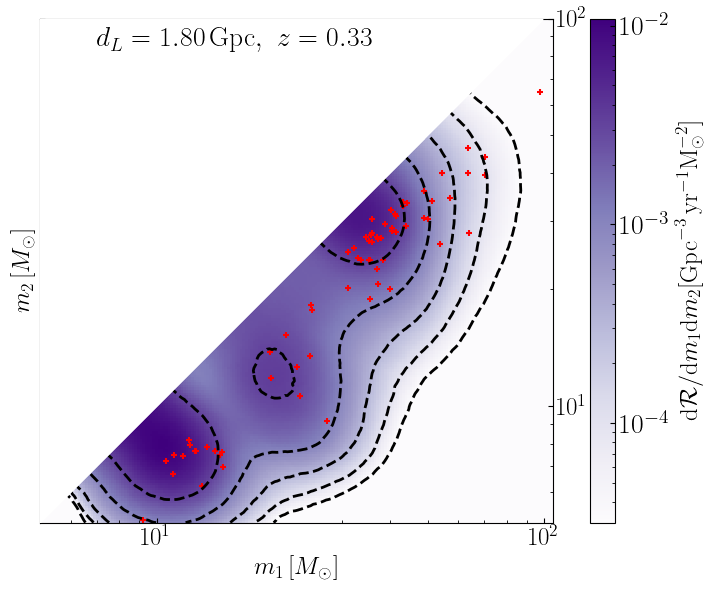}
\caption{Rate density as a function of component masses at fixed luminosity distance or redshift values.  Red $+$ markers represent the medians of PE samples.
}
\label{fig:3D-m1-m2fixed-dL}
\end{figure}
We also select different $d_L$ values to plot the median rate over component masses, as in Fig.~\ref{fig:3D-m1-m2fixed-dL} (note the different color scales between the two panels).  The distribution at $0.9\,$Gpc shown in the upper plot is essentially unchanging for distances out to the detection horizon for $10\,M_\odot$ components; the lower plot suggests a shift towards apparent higher support for high BH masses at larger distances, however the principal change here is again a suppression of the low-mass component due to very small \pdet\ (censorship). 

\begin{figure}[tbp]
    \centering
    \includegraphics[width=0.93\linewidth]{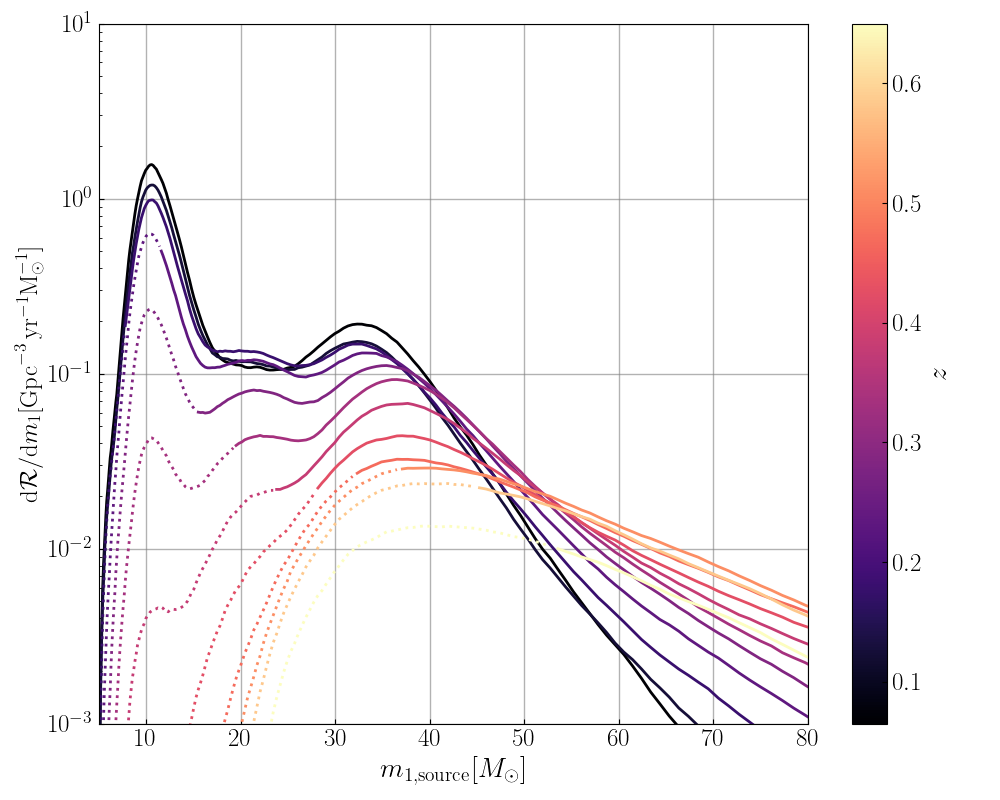}
    \caption{Merger rate density (median estimate) as a function of the primary mass $m_1$, integrated over secondary mass, with different curves representing rates at redshift values indicated by color.
    Regions with $\pdet < 0.1$, estimated as in Fig.~\ref{fig:median-m1-rate-powerlaw-m2}, are shown as dotted curves.}
    \label{fig:3Dmedian-rate}
\end{figure}
\begin{figure}[tbp]
    \centering
    \includegraphics[width=0.99\linewidth]{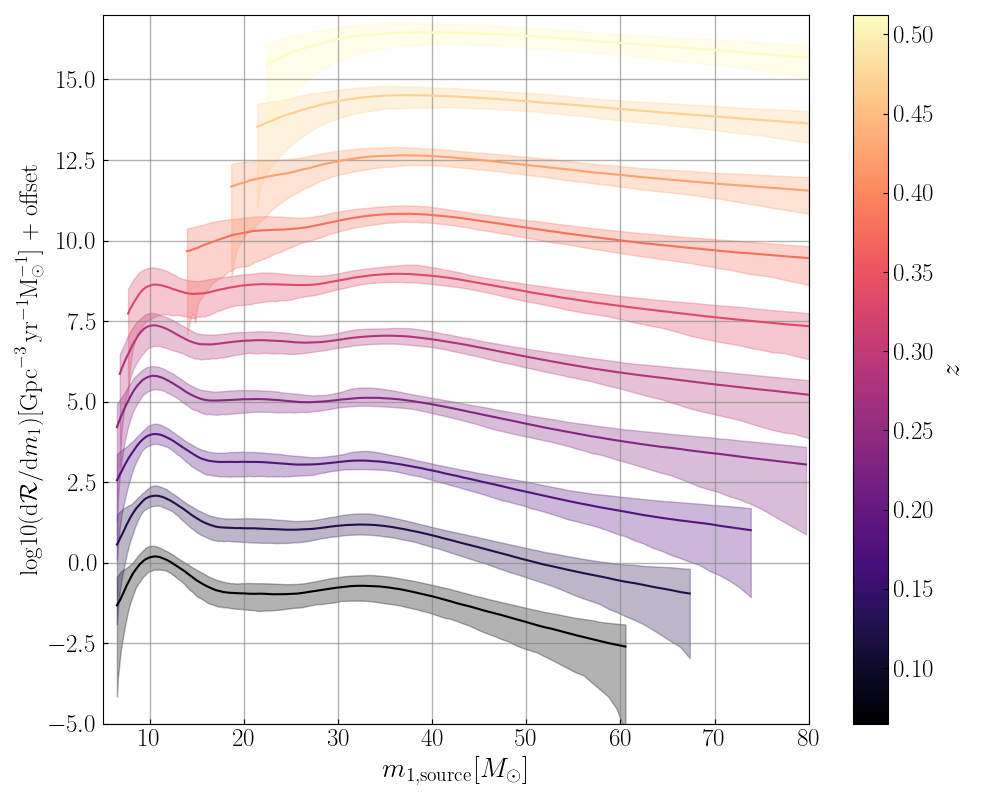}
    \caption{Merger rate density as a function of the primary mass $m_1$, corresponding to the results shown in Fig.~\ref{fig:3Dmedian-rate}. 
    Curves are defined as in Fig.~\ref{fig:offset-m1-rate-powerlaw-m2}. 
    }
    \label{fig:3Doffset-rate}
\end{figure}

We further investigate potential evolution by taking several fixed $d_L$ (or $z$) values and integrating the joint rate over $m_2$ to obtain rate vs.\ primary mass $m_1$.  Median rates are shown in Fig.~\ref{fig:3Dmedian-rate} and rates with uncertainties in Fig.~\ref{fig:3Doffset-rate}, showing similar trends to the 2D case where a power-law assumption was applied to $m_2$. 
Fig.~\ref{fig:3Dmedian-rate} suggests a more visible trend towards shallower fall-off of the high mass distribution with increasing redshift; however, uncertainties remain extremely large for $m_1>50\,M_\odot$.  Hence, again we do not see positive evidence for evolution.  

While this conclusion agrees with most other studies, it is in contrast to \cite{Rinaldi:2023bbd} which claims non-trivial evolution, specifically decreasing low-mass rates, resp.\ increasing higher-mass rates with redshift.  While the present work shares a method of estimating selection effects with \cite{Rinaldi:2023bbd}, i.e.\ the injection fit of~\cite{Lorenzo-Medina:2024opt}, our inference on the detected BBH distribution is obtained with a completely different and independent analysis.  We observe similar trends, however the large uncertainties we obtain over much of the parameter space imply the trends are not significant.

Our alternative analysis using sample $\chi_\mathrm{eff}$ values to calculate $\pdet$ for the reweighting steps produces results, shown in Appendix~\ref{app:xieffpdet}, extremely similar to our main analysis.  The analysis with unrestricted $d_L$ bandwidth choice described in Appendix~\ref{app:bwdLsmaller}, which might in principle be sensitive to rapid changes in rate over redshift, does not show features beyond those already noted, while having substantially larger uncertainties attributable to random sample fluctuations.  

We also conduct an verification analysis imposing a stricter bound on FAR of $0.25/$y to check for possible bias due to noise contamination of our event set~\footnote{We thank the referee for suggesting this check.}: as shown in Appendix~\ref{app:stricter_far}, we find no significant changes in the overall inference, although statistical uncertainties increase due to the lower detection count. 
 
One further common feature of our rate estimates is that they slightly decrease with redshift, even well within the ``horizon'' of detected events where $\pdet\gtrsim 0.1$, except for a possible increasing trend up to $z \sim 0.5$ at high masses.  This is in apparent contrast to some models of redshift evolution, for instance, taking the comoving rate to vary proportional to $\mathcal{R}_0(1+z)^\kappa$ a generally positive, though small value of $\kappa$ is inferred~\cite{KAGRA:2021duu}.  However, non-parametric approaches to measuring evolution of the total BBH merger rate \emph{do not confirm} an increasing rate at small redshifts $\lesssim 0.5$, some even showing a slight decrease \cite{Edelman:2022ydv,Callister:2023tgi,Ray:2023upk}.  Only for $z\gtrsim 0.6$ is an increasing trend visible in such studies -- though not in our results.  As seen e.g.\ in Fig.~\ref{fig:m1dL2D_withselectioneffectContours}, high-redshift estimates must be driven almost entirely by high-mass events. 

Such apparent discrepancies may require further investigation, as finite, though hopefully small, biases can arise, either in our method or in (semi-)parameterized models. 
On the KDE side the choice of a constant kernel (up to local adaptive scaling) in the logarithm of masses and in luminosity distance may cause some bias given the current low number of detections; we expect such biases to be reduced with higher event counts, as optimal bandwidths will become smaller and the exact choice of kernel becomes less important.  Conversely, parameterized evolution models, or even those allowing for more general $z$-dependence, typically assume specific functional forms for many parameters such as mass ratio and spins.  Even when dependence on every parameter is described non-parametrically~\cite{Edelman:2022ydv}, any assumption that different parameters are uncorrelated is a significant restriction.  Such assumptions may lead to biases in estimating the selection function or in individual event parameters, which may partially mimic an evolving rate, since rate estimates at different redshifts are informed by events with different intrinsic properties.  Our current analysis also does not allow for the BBH effective spin distribution to evolve over redshift; nontrivial evolution, as suggested in \cite{Biscoveanu:2022qac}, could impact inferences at high $z$ via the selection function.

\section{Discussion}

\label{sec:discussion}
The detection of mass evolution with redshift using gravitational waves would have significant astrophysical implications, offering insights into the formation and evolution of compact binary systems over cosmic time and potentially helping to untangle different formation channels and environments. Recent studies using non-parametric or parametrized change-point methods with the same input data come to different conclusions. We use our non-parametric, data driven method, accounting for selection effects, to investigate this question and found no clear evidence, within the current limitations of observation selection effects and finite (low number) statistics.

The immediate implications from this study in are methodological rather than astrophysical.  Given the low statistics of detections at masses above $\sim\!35\,M_\odot$ and non-detectability of low-mass BBH at redshifts above $z\sim 0.2$ in GWTC-3, our ability to infer \emph{anything} about evolution of the BH mass spectrum is extremely limited.  Apparent disagreements in previous inferences are probably due to different implicit assumptions about the behaviour of the population, inherent in the different methods employed.  On the non-parametric side, methods may be prone to over-fit random fluctuations in observed event parameters, or underestimate uncertainties in regions with few (or zero!) detections.  On the (semi-)parametric side, possible biases due to inaccurate model assumptions, for example on mass ratio and spins, are difficult to quantify directly.  With increasing detector sensitivities and detection counts, different methods should converge, however meaningful probes of evolution may still be some way in the future.  In particular, probing the high-mass end of the population will simply require significantly longer observing times: existing detectors are already sensitive out to moderately large redshift but statistics remain low due to the sheer rarity of mergers. 

Overall, detecting mass evolution with redshift would provide a powerful tool to bridge gravitational-wave observations with broader astrophysical and cosmological phenomena. 
In future we will apply this method to the  significantly larger set of observations expected from O4 and future runs, and further explore correlations with other parameters, such as spins.  

\section*{Acknowledgements}
We thank the LVK compact binary Rates \& Populations working group for useful discussions. 
J.S.\ acknowledges support from the European Union’s H2020 ERC Consolidator Grant ``GRavity from Astrophysical to Microscopic Scales'' (Grant No. GRAMS-815673), the PRIN 2022 grant ``GUVIRP - Gravity tests in the UltraViolet and InfraRed with Pulsar timing'', and the EU Horizon 2020 Research and Innovation Programme under the Marie Sklodowska-Curie Grant Agreement No. 101007855. T.D.\ 
has received financial support from Xunta de Galicia (CIGUS Network of research centers) and the European Union. 

The authors are grateful for computational resources provided by the LIGO Laboratory and supported by National Science Foundation Grants PHY-0757058 and PHY-0823459. This research has made use of data or software obtained from the Gravitational Wave Open Science Center (gwosc.org), a service of the LIGO Scientific Collaboration, the Virgo Collaboration, and KAGRA. This material is based upon work supported by NSF's LIGO Laboratory which is a major facility fully funded by the National Science Foundation, as well as the Science and Technology Facilities Council (STFC) of the United Kingdom, the Max-Planck-Society (MPS), and the State of Niedersachsen/Germany for support of the construction of Advanced LIGO and construction and operation of the GEO600 detector. Additional support for Advanced LIGO was provided by the Australian Research Council. Virgo is funded, through the European Gravitational Observatory (EGO), by the French Centre National de Recherche Scientifique (CNRS), the Italian Istituto Nazionale di Fisica Nucleare (INFN) and the Dutch Nikhef, with contributions by institutions from Belgium, Germany, Greece, Hungary, Ireland, Japan, Monaco, Poland, Portugal, Spain. KAGRA is supported by Ministry of Education, Culture, Sports, Science and Technology (MEXT), Japan Society for the Promotion of Science (JSPS) in Japan; National Research Foundation (NRF) and Ministry of Science and ICT (MSIT) in Korea; Academia Sinica (AS) and National Science and Technology Council (NSTC) in Taiwan.

\section*{DATA AVAILABILITY}
The data that support the findings of this article are openly available at \cite{ligo_scientific_collaboration_and_virgo_2021_5546663} and associated analysis data are available at \url{https://zenodo.org/records/17178478}. 

\newpage
\bibliography{reference}
\clearpage 

\appendix

\section{Spin dependence of selection function in reweighting}
\label{app:xieffpdet}

\begin{figure}[tbp]
    \centering
    \includegraphics[width=0.9\linewidth]{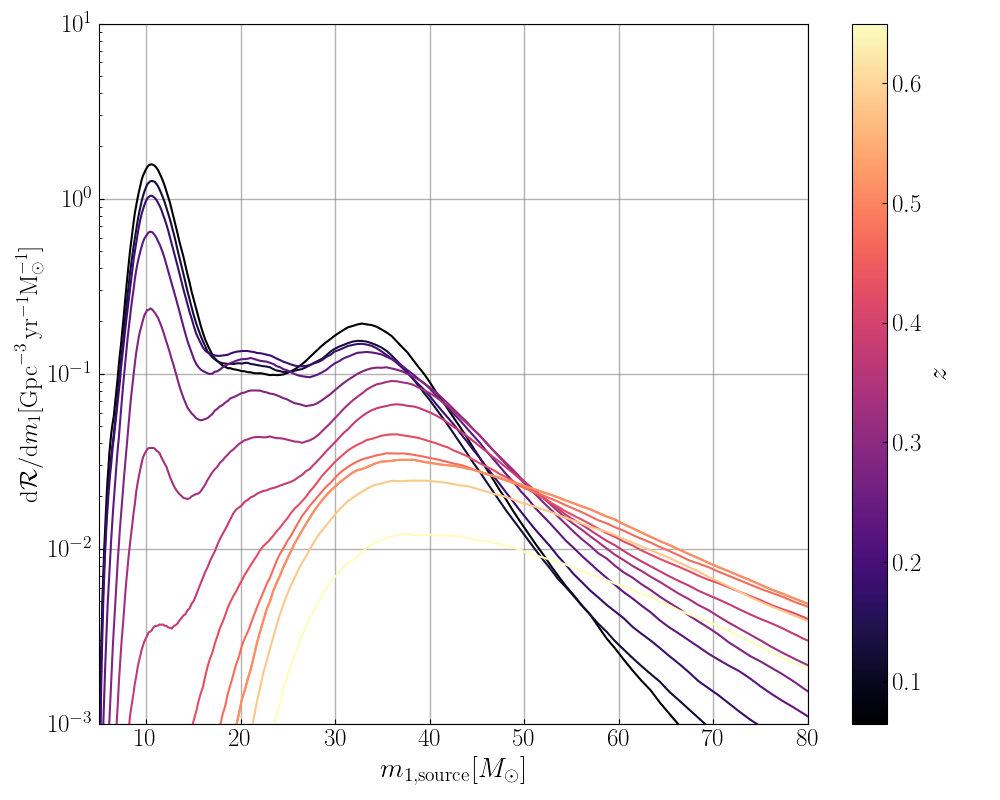}
    \caption{Alternative analysis using sample $\chi_\mathrm{eff}$ values to evaluate $\pdet$.  Curves defined as in Fig.~\ref{fig:3Dmedianbw001}.
    }
    \label{fig:Xieffbased_3Dmedianbw001}
\end{figure}

In our reweighting procedure, we incorporate selection effects into the parameter estimation samples. However, in our main study, these selection effects are estimated with $\chi_\mathrm{eff} = 0$. 
\begin{figure}[tbp]
    \includegraphics[width=0.85\linewidth]{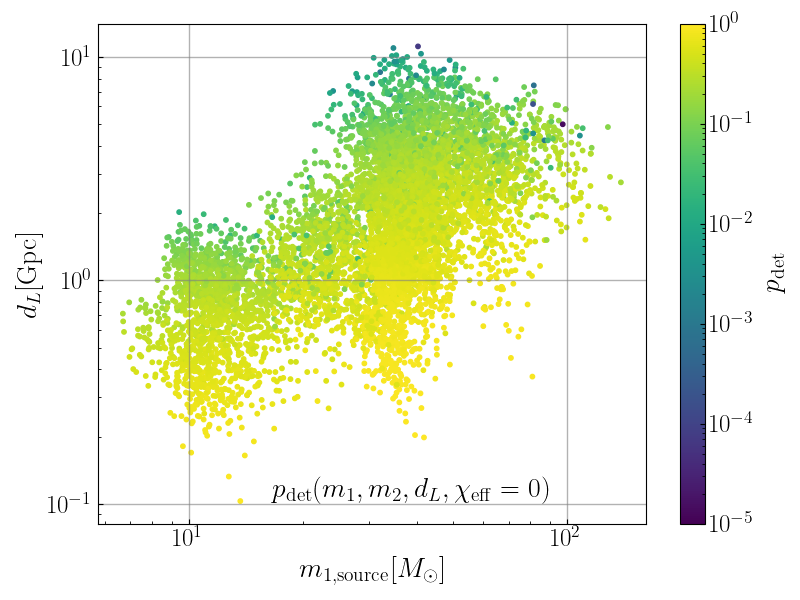}
    \includegraphics[width=0.85\linewidth]{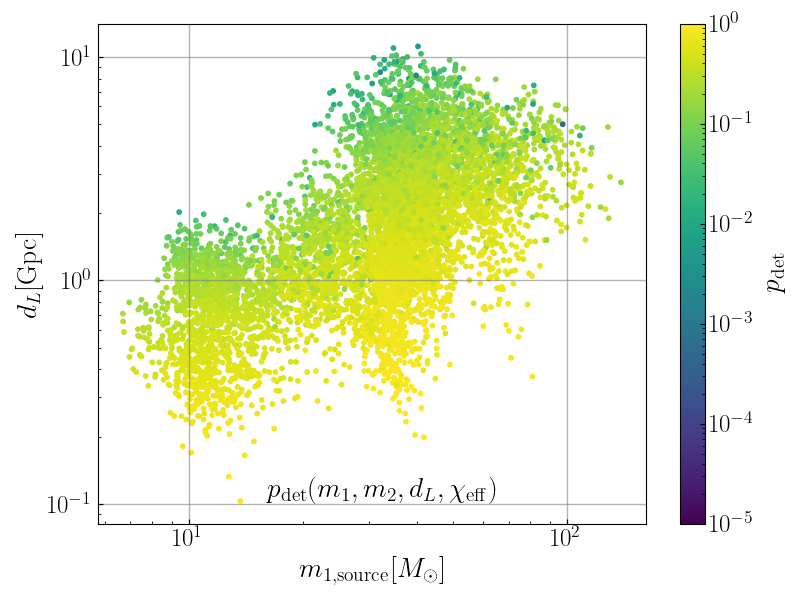}
    \caption{Detection probability $\pdet$ as a function of the primary mass $m_1$ and luminosity distance $d_L$ for GWTC-3 parameter estimation samples. 
    Top: setting $\chi_\mathrm{eff}=0$ in evaluation of $\pdet$ (as in our main analysis). Bottom: using sample $\chi_\mathrm{eff}$ values in evaluation of $\pdet$. 
    }
    \label{fig:Xieffbasedpdet-on-pesamples}
\end{figure}
Here, we investigate possible spin effects by using the PE sample $\chi_\mathrm{eff}$ values in evaluating $\pdet$ during the iterative reweighting procedure.  We see generally small differences, as shown by the comparison in Fig.~\ref{fig:Xieffbasedpdet-on-pesamples}. 
In this analysis, we follow the same procedure as Sec.~\ref{ssec:pdet} capping the maximum value of $\pdet$ at $0.1$ in order to regularize behaviour in regions with very small $\pdet$,  
and enforce a minimum bandwidth of $0.3$ over $d_L$. 
The resulting median rate estimates at fixed distance/redshift values shown in Fig.~\ref{fig:Xieffbased_3Dmedianbw001} are extremely similar to the main results which set $\chi_\mathrm{eff}=0$ in reweighting. 

\section{Results with unrestricted bandwidth over distance}
\label{app:bwdLsmaller}

\begin{figure}
\includegraphics[width=0.9\linewidth]{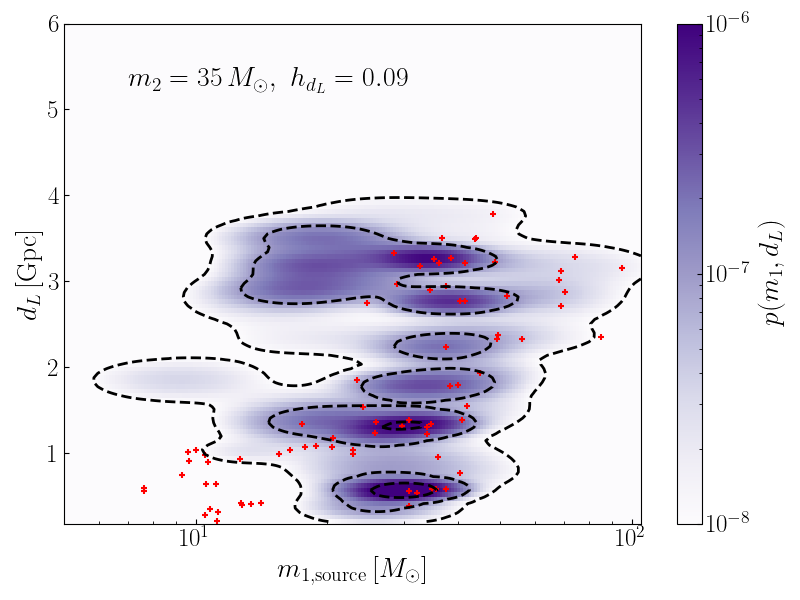}
    \caption{Density estimate for a single iteration with small optimal $d_L$ bandwidth, at a fixed $m_2$ slice, showing rapid variation in rate over distance/redshift. 
    Red crosses indicate the samples selected in this iteration during the reweighting process.}
\label{fig:bwdLissue}
\end{figure}
\begin{figure}[tbp]
    \centering
    \includegraphics[width=0.85\linewidth]{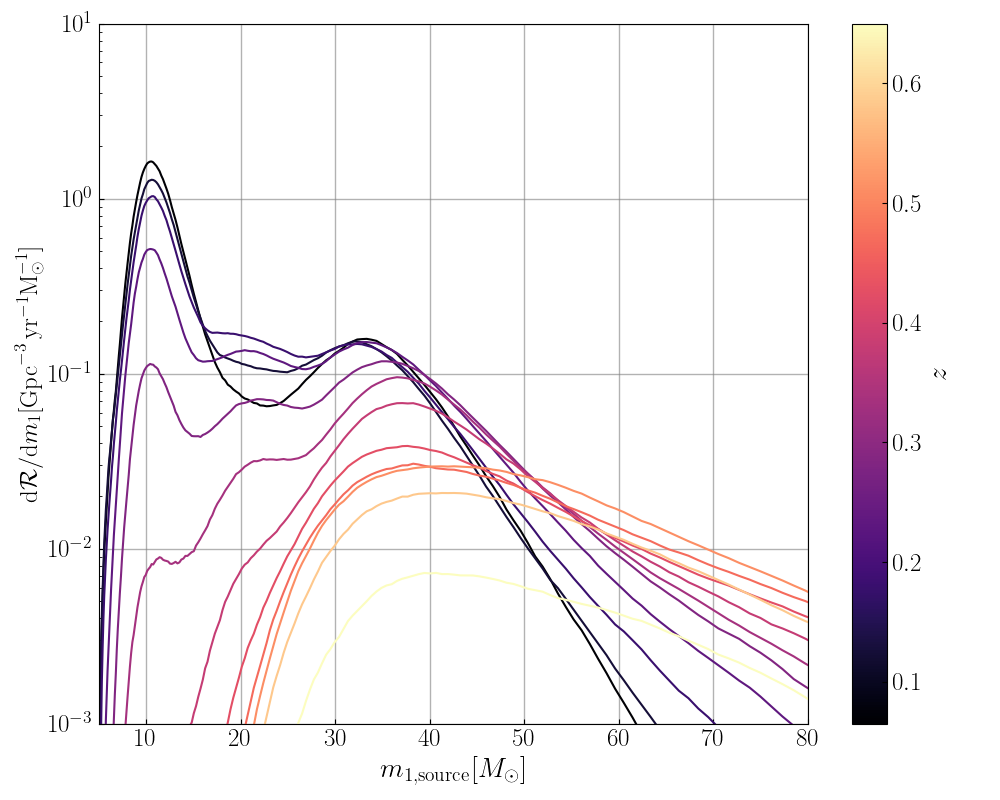}
    \caption{
    Merger rate density (median estimate) as a function of the primary mass $m_1$, integrated over secondary mass, with different curves representing rates at redshift values indicated by color. 
    This analysis uses unrestricted bandwidth over $d_L$, in contrast to our main result. 
    }
    \label{fig:3Dmedianbw001}
\end{figure}
\begin{figure}[tbp]
    \centering
    \includegraphics[width=0.9\linewidth]{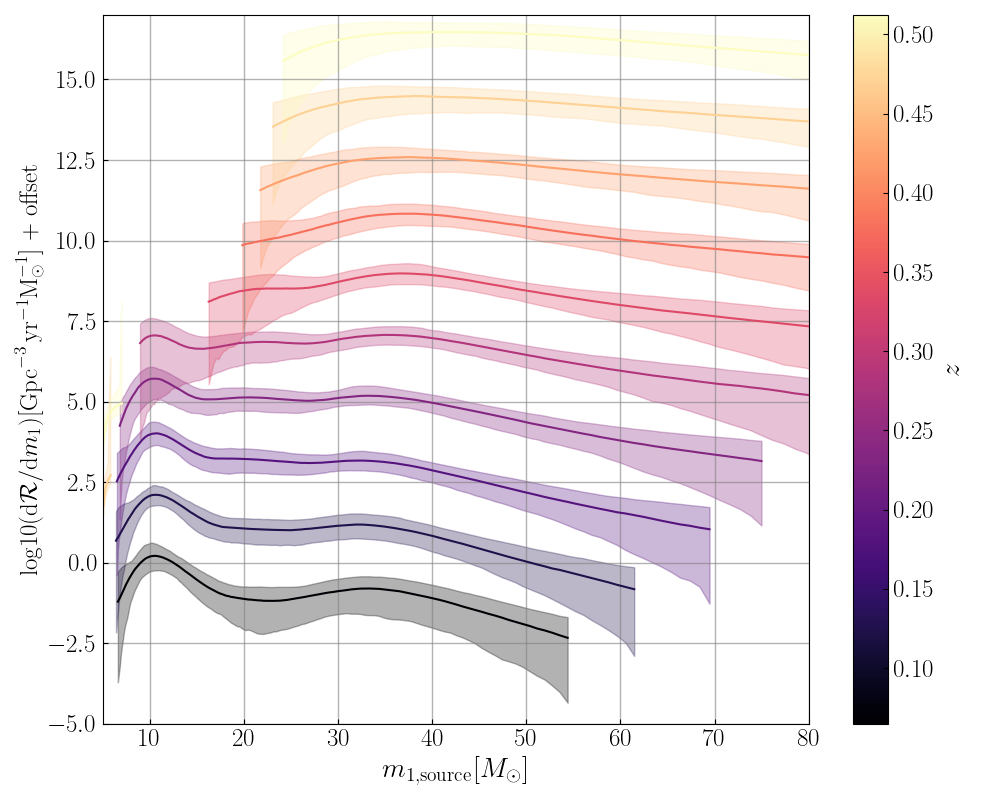}
    \caption{Merger rate density as a function of the primary mass $m_1$ with uncertainties, corresponding to the results shown in Fig.~\ref{fig:3Dmedianbw001}. 
    Curves are defined as in Fig.~\ref{fig:offset-m1-rate-powerlaw-m2}. 
    }
    \label{fig:3Doffsetbw001}
\end{figure}

We perform the same 3D analysis as described in the main text, but without imposing any lower limit on the optimized bandwidth in the $d_L$ dimension. Out of 1000 iterations, a few have an optimal bandwidth smaller than 0.1: Fig.~\ref{fig:bwdLissue} illustrates the effect of a small $d_L$ bandwidth on the KDE, showing a physically unrealistic strong multimodal variation over redshift. 
Overall, the results over all iterations remain largely unchanged, though with higher uncertainties, as illustrated in Figs.~\ref{fig:3Dmedianbw001} and \ref{fig:3Doffsetbw001}.

\section{Conversion to Cosmological Rate Density}
\label{app:comov_rate}

As described in \ref{sec:results}, we derive a KDE rate density, i.e.\ number density per observation time, over intrinsic parameters $\theta$ (here component mass) and luminosity distance, from the normalized KDE $p(\theta, d_L)$ as
\begin{equation}
 \frac{\D N}{\D \theta \D d_L \D t_\mathrm{det}} = \frac{p(\theta, d_L) N_\mathrm{obs}}{\pdet(\theta, d_L) T_\mathrm{obs}}.
\end{equation}
We convert this to astrophysical rate density per comoving volume per source-frame time, $\D \mathcal{R}/\D \theta \equiv \D N/(\D \theta \D V_c \D t_\mathrm{s})$
via 
\begin{equation}
 \D \mathcal R = \frac{\D N}{\D d_L \D t_\mathrm{det}}
 \frac{\D t_\mathrm{det}}{\D t_\mathrm{s}} 
 \frac{\D d_L}{\D z}
 \left(\frac{\D V_c}{\D z}\right)^{-1}.\vspace*{0.05cm}
\end{equation}
We have $\D t_\mathrm{det}/\D t_\mathrm{s} = 1 + z$ due to time dilation.  Further, for flat $\Lambda$CDM cosmology we have (e.g.~\cite{Hogg:1999ad})
\begin{equation}
 \frac{\D V_c}{\D z} = 4\pi d_c^2 \frac{\D d_c}{\D z}
 = \frac{c}{H_0}\frac{4\pi d_c^2}{E(z)}\,,
\end{equation}
where $d_c$ is the comoving line-of-sight distance
\begin{equation}
 d_c = \frac{c}{H_0} \int_0^z \frac{dz'}{E(z')},\quad
 E(z) = \sqrt{\Omega_m (1+z)^3 + \Omega_\Lambda}\,,
\end{equation}
thus for the luminosity distance we have
\begin{equation}
 d_L = (1 + z) d_c \implies \frac{\D d_L}{\D z} = 
 d_c + \frac{c}{H_0}\frac{(1 + z)}{E(z)}. 
\end{equation}

\section{Analysis of Events With FAR $\leq 0.25/$y}
\label{app:stricter_far}

Motivated by the possibility of noise contamination in our event set, we apply a stricter cut on the FAR in the selection of BBH events to examine its impact on the results. With this threshold set to $0.25$ per year, we obtain a total of 62 BBH events from the GWTC-3 catalog (we also recalculate \pdet\ corresponding to this more stringent threshold). Here we present results derived from these selected events.

\begin{figure}[tbp]
    \centering
    \includegraphics[width=0.9\linewidth]{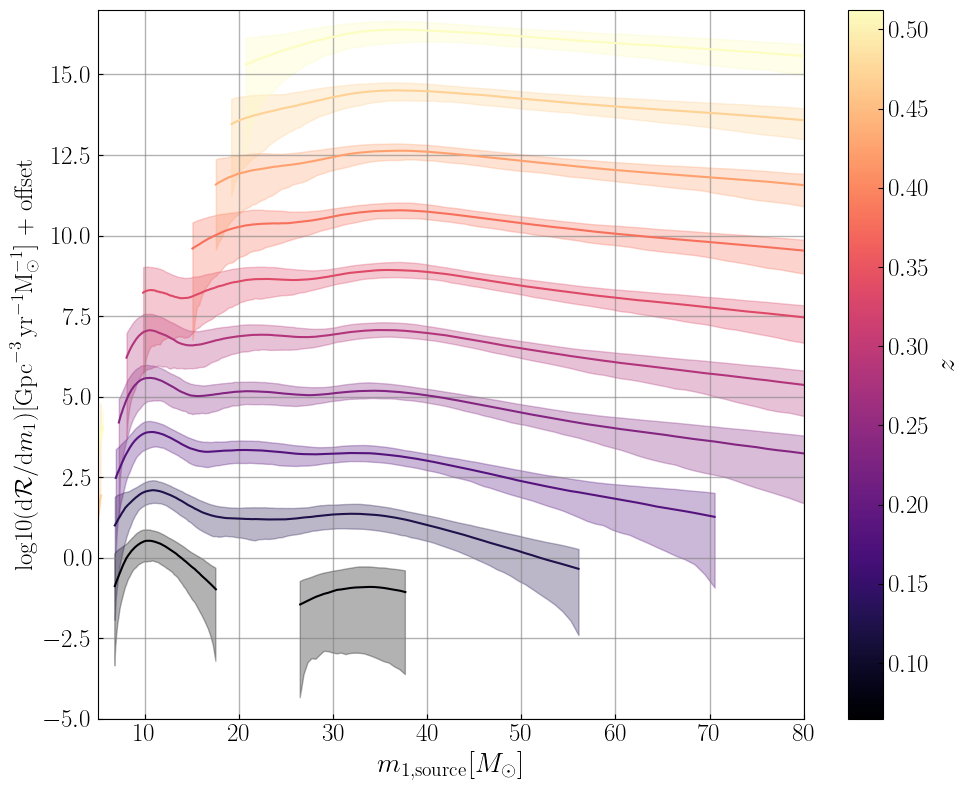}
    
    \caption{Merger rate density as a function of the primary mass $m_1$ with uncertainties, where only events with FAR $\leq$ 0.25 are included (62 BBH events).}
    \label{fig:3Doffsetwith62BBHeventsFAR025}
\end{figure}

As shown in Figure~\ref{fig:3Doffsetwith62BBHeventsFAR025}, 
we obtain similar results to our main analysis.  However, we see much larger uncertainties for $m_1 \gtrsim 15\,M_\odot$ at low redshifts.  Such differences may be driven by a preference for different KDE bandwidth and adaptive parameter values as compared to our main analysis, given a smaller number of detections occupying a narrower overall range of $d_L$.

\end{document}